\documentstyle{article}

\font\tenrm=cmr10
\font\tenit=cmti10
\font\elevenbf=cmbx10 scaled\magstep 1
\font\elevenrm=cmr10 scaled\magstep 1
\font\elevenit=cmti10 scaled\magstep 1

\renewenvironment{thebibliography}[1]
 { \elevenrm
   \begin{list}{\arabic{enumi}.}
    {\usecounter{enumi} \setlength{\parsep}{0pt}
     \setlength{\itemsep}{3pt} \settowidth{\labelwidth}{#1.}
     \sloppy
    }}{\end{list}}

\begin{document}
\hfill SHEP 92/93-19
\begin{center}
\vglue 0.6cm
{\elevenbf
\vglue 10pt BOUNDING THE HIGGS BOSON MASS IN THE\\
\vglue 3pt  NEXT--TO--MINIMAL SUPERSYMMETRIC STANDARD MODEL
\footnote{Contribution to SUSY--93 Proceedings, International
Workshop on Supersymmetry and Unification of Fundamental
Interactions, Boston, U.S.A., March 29 -- April 1, 1993.}\\}
\vglue 1.0cm
{\tenrm T. ELLIOTT \\}
\baselineskip=13pt
{\tenit Physics Department, University of Southampton,
Highfield,\\}
\baselineskip=12pt
{\tenit Southampton, SO9 5NH, England.\\}
\vglue 0.8cm
{\tenrm ABSTRACT}
\end{center}
\vglue 0.3cm
{\rightskip=3pc
 \leftskip=3pc
 \tenrm\baselineskip=12pt
 \noindent
We discuss the upper bound on the lightest CP-even Higgs boson
mass in the
next--to--minimal supersymmetric standard model within the
framework of a
low energy renormalisation group analysis. We find
$m_h < 146$ GeV for
$m_t = 90$ GeV, decreasing to $m_h < 123$ GeV for
$m_t=180$ GeV.}

\vglue 0.6cm
\vglue 0.4cm
\baselineskip=14pt
\elevenrm
The minimal supersymmetric standard model \cite{nilles} (MSSM) is
the simplest supersymmetric (SUSY) generalisation of the standard
model which has many attractive features.  Because of the
non-renormalisation theorem and phenomenological SUSY, the
technical hierarchy problem is solved, namely, why is $M_Z \ll
M_{planck}$ stable against perturbation theory? Moreover, the
realisation that the electroweak group, $SU(2)_L \otimes U(1)_Y$,
may be broken radiatively due to the top quark's large Yukawa
coupling provides an elegant and perhaps compelling explanation
for the smallness of $M_Z$ compared to $M_{planck}$. Furthermore,
it is easy to bound the lightest CP-even Higgs boson mass at
tree--level, yielding the classic result $m_h < M_Z$, though it
is well--known that the tree--level bound is subject to large
radiative corrections, the size of which may be estimated using
triviality limits on the top quark Yukawa coupling. Radiative
corrections to the Higgs boson masses in the MSSM have been the
subject of much recent discussion \cite{mssmbound}.  However, one
aspect of the MSSM which is unsatisfactory is the so--called
$\mu$--problem, the occurrence of a mass--scale $\mu \sim O(M_Z)$
in the superpotential. One would naively expect, in a SUSY
theory, that either $\mu \equiv 0$ or $\mu \sim O(M_{planck})$.

The MSSM is not, however, the only generalisation of the standard
model compatible with grand unification. It is possible that SUSY
grand unified theories (GUTs) give rise to a low energy theory
containing an additional gauge singlet field, the so--called
next--to--minimal supersymmetric standard model
\cite{nmssmdefined,treeboundbib} (NMSSM). In this talk we shall
be concerned with bounding the lightest CP-even Higgs mass in
this more general model by using the triviality limits of Yukawa
couplings in the theory.

The NMSSM contains two Higgs doublets and one Higgs gauge singlet
field, leading to three CP-even neutral states, two CP-odd
neutral states and two charged bosons in the physical spectrum,
and is defined by the superpotential
\begin{equation}
W = h_UQH_2U^c + h_DQH_1D^c + \lambda NH_1H_2 - \frac{1}{3} kN^3,
\end{equation}
where gauge and family indices are suppressed, $H_{1,2}$ contain
the standard Higgs doublets and $N$ contains the Higgs gauge
singlet. The only Yukawa couplings which we retain are $h_t$, the
top quark Yukawa coupling, $\lambda$ and $k$, since these may all
be potentially large (we assume $h_b \ll h_t$). Thus, the
superpotential becomes
\begin{equation}
W = h_tQH_2t^c + \lambda NH_1H_2 - \frac{1}{3} kN^3,
\end{equation}
where $Q^T = (t_L,b_L)$ contains the left--handed top and bottom
quarks, and $t^c$ contains the charge conjugate of the
right--handed top quark.

The $\mu$--problem is eliminated in the NMSSM by replacing $\mu$
by the gauge singlet $N$ which develops a vacuum expectation
value (vev), $<N>=x$. The remaining vevs are
$<H_{1,2}>={\nu}_{1,2}$, where
$\nu=\sqrt{{\nu}_1^2+{\nu}_2^2}=174$ GeV. The trilinear term
$\frac{1}{3} kN^3$ removes a global $U(1)$ symmetry which would
result in an unwanted axion when the fields acquire their vevs.

An upper bound on the lightest CP-even scalar $h^0$ in the NMSSM
may be obtained from the real, symmetric $3 \times 3$ CP-even
scalar mass--squared matrix, by using the fact that its minimum
eigenvalue is bounded by the minimum eigenvalue of its upper $2
\times 2$ submatrix. In this manner, we may obtain the
tree--level bound \cite{treeboundbib}
\begin{equation}
\label{treebound}
m_h^2 \leq M_Z^2 + ({\lambda}^2{\nu}^2 - M_Z^2) \sin^2 2\beta,
\end{equation}
where $\tan \beta = {\nu}_2 / {\nu}_1$, and $\lambda = \lambda
(M_{SUSY})$.  This is maximised by taking $\lambda$ as large as
possible; the triviality limit, $\lambda = {\lambda}_{max}$,
provides the most reasonable maximum value, for a given top quark
mass. ${\lambda}_{max}$ is determined by solving the SUSY
renormalisation group (RG) equations for the Yukawa couplings
$h_t$, $\lambda$, and $k$ in the region $M_{SUSY} = 1$ TeV to
$M_{GUT} = 10^{16}$ GeV \cite{triviality}.

Radiative corrections to Eq.~(\ref{treebound}) have recently been
considered using either the Effective Potential method
\cite{epbound}, or the RG approach \cite{rgbound}. However, the
implementations of the latter have assumed either SUSY RG
equations down to near the top mass, or the existence of only one
light Higgs particle in the low energy spectrum below $M_{SUSY}$.
Here we shall perform a low energy RG analysis of the Higgs
sector of the model between $M_{SUSY}$ and some lower scale
$\mu$, assuming that all Higgs scalars may have masses less than
$M_{SUSY}$, and using non--SUSY RG equations below $M_{SUSY}$. We
make the usual approximation of hard decoupling of superpartners
at $M_{SUSY}$.  We then apply this technique to obtain a
radiatively corrected upper bound on $m_h$.

The general low energy scalar Higgs potential is
\begin{eqnarray}
V_{Higgs} & = & \frac{1}{2} {\lambda}_1(H_1^{\dagger} H_1)^2 +
\frac{1}{2} {\lambda}_2(H_2^{\dagger} H_2)^2 + ({\lambda}_3
+{\lambda}_4)(H_1^{\dagger} H_1)(H_2^{\dagger} H_2) \nonumber \\
& - & {\lambda}_4|H_2^{\dagger} H_1|^2 + {\lambda}_5|N|^2|H_1|^2
+ {\lambda}_6|N|^2|H_2|^2 \nonumber \\ & + & {\lambda}_7(N^{\ast
2} H_1H_2 + h.c.)  + {\lambda}_8|N|^4 \nonumber \\ & + &
m_1^2|H_1|^2 + m_2^2|H_2|^2 + m_3^2|N|^2 \nonumber \\ & - &
m_4(H_1H_2N + h.c.)- \frac{1}{3} m_5(N^3 + h.c.),
\end{eqnarray}
and we have the following boundary conditions at $M_{SUSY}$:
\[
{\lambda}_1 = {\lambda}_2=\frac{1}{4} (g_2^2 + g_1^2),\ \ \
{\lambda}_3=\frac{1}{4} (g_2^2 - g_1^2),
\]
\[
{\lambda}_4 = {\lambda}^2-\frac{1}{2}g_2^2,\ \ \
{\lambda}_5={\lambda}_6={\lambda}^2, \ \ \
{\lambda}_7=-{\lambda}k, \ \ \ {\lambda}_8=k^2,
\]
\begin{equation}
m_1 = m_{H_1}, \ \ \ m_2 = m_{H_2}, \ \ \ m_3 = m_N, \ \ \ m_4 =
{\lambda}A_{\lambda}, \ \ \ m_5=kA_k.
\label{bc}
\end{equation}
$g_1$ and $g_2$ are the $U(1)_Y$ and $SU(2)_L$ gauge coupling
constants, at $M_{SUSY}$, respectively, and $M_{H_i}$, $M_N$,
$A_{\lambda}$ and $A_k$ are soft SUSY breaking parameters. We do
not consider the effects of any other soft SUSY breaking
parameters. The low energy couplings ${\lambda}_i(\mu)$ may be
obtained by solving their RG equations between $M_{SUSY}$ and
$\mu$. We do not reproduce the RG equations here, but they may be
found elsewhere \cite{elliott}. In solving these equations, we
decouple the top quark at its mass shell, but do not decouple the
Higgs scalars at their mass shells. We do not expect the latter
to be a significant approximation provided $\mu \sim m_h$, and
the Higgs scalars are light.

The minimisation conditions implied by $\frac{ \partial
V_{Higgs}}{\partial v_i}=0$ and $\frac{ \partial
V_{Higgs}}{\partial x}=0$ allow us to eliminate the low energy
parameters $m_1$, $m_2$ and $m_3$ in favour of ${\nu}_1$,
${\nu}_2$ and $x$. The remaining parameters $m_4$ and $m_5$
cannot so be eliminated, but we may remove $m_4$ in favour of the
charged Higgs scalar mass, $m_c$, by using the result
\begin{equation}
m_c^2 = \frac{2x}{\sin 2\beta} (m_4 - {\lambda}_7 x) -
{\lambda}_4 {\nu}^2.
\end{equation}
Thus, we are then left with the following parameters in our low
energy theory: $\frac{{\nu}_2}{{\nu}_1} = \tan \beta$,
$\frac{x}{\nu} = r$, $\lambda$, $k$, $m_5$, $m_c$.

The upper $2 \times 2$ submatrix of the full $3 \times 3$ CP-even
mass--squared matrix (in the basis $\{H_1,H_2,N\}$) is given by
\begin{equation}
M^2 = \left( \begin{array}{cc} 2{\lambda}_1 {\nu}_1^2 &
2({\lambda}_3+{\lambda}_4){\nu}_1{\nu}_2 \\
2({\lambda}_3+{\lambda}_4){\nu}_1{\nu}_2 & 2{\lambda}_2 {\nu}_2^2
\end{array} \right) + \left( \begin{array}{cc} \tan \beta & -1 \\
-1 & \cot \beta \end{array} \right) \frac{1}{2} (m_c^2 +
{\lambda}_4{\nu}^2) \sin 2\beta .
\end{equation}
Notice that this is independent of both $m_5$ and $r$. From this
we see that
\begin{equation}
m_h^2 \leq
\frac{1}{2}(A+m_c^2)-\frac{1}{2}\sqrt{(m_c^2+B)^2+C^2-B^2},
\end{equation}
where
\begin{eqnarray}
A & = & {\nu}^2({\lambda}_1+{\lambda}_2+{\lambda}_4) +
{\nu}^2({\lambda}_1-{\lambda}_2)\cos 2\beta , \nonumber \\ -B & =
& {\nu}^2 \left[ ({\lambda}_1-{\lambda}_2) +
({\lambda}_1+{\lambda}_2-{\lambda}_4)\cos 2\beta \right] \cos
2\beta + {\nu}^2(2{\lambda}_3+{\lambda}_4)\sin^2 2\beta ,
\nonumber \\ C^2 & = & {\nu}^4 \left[ ({\lambda}_1-{\lambda}_2) +
({\lambda}_1+{\lambda}_2-{\lambda}_4)\cos 2\beta \right]^2 +
{\nu}^4 (2{\lambda}_3+{\lambda}_4)^2 \sin^2 2\beta .
\label{ABC}
\end{eqnarray}
Since $C^2-B^2 \geq 0$, we obtain the $m_c$--independent bound
\begin{equation}
m_h^2 \leq \frac{1}{2} (A-B) .
\end{equation}
Inserting $A$ and $B$ from Eq.~(\ref{ABC}), and using the
boundary conditions in Eq.~(\ref{bc}) leads to an upper bound of
the form
\begin{eqnarray}
m_h^2 & \leq & M_Z^2 + (\lambda^2 {\nu}^2 - M_Z^2) \sin^2 2\beta
\nonumber \\ & + & \frac{{\nu}^2}{2}[(\delta {\lambda}_1 + \delta
{\lambda}_2 + 2\delta {\lambda}_3 + 2\delta {\lambda}_4) +
2(\delta {\lambda}_1 - \delta {\lambda}_2)\cos 2\beta \nonumber
\\ & + & (\delta {\lambda}_1 + \delta {\lambda}_2 - 2\delta
{\lambda}_3 - 2\delta {\lambda}_4) \cos^2 2\beta],
\label{analyticbound}
\end{eqnarray}
where $\delta {\lambda}_i = {\lambda}_i(\mu) -
{\lambda}_i(M_{SUSY})$, and, strictly, $M_Z=M_Z(M_{SUSY})$. By
approximating the RG equations (retaining terms quadrilinear in
$\lambda$, $h_t$ and $k$ only, and making a small $\delta
{\lambda}_i$ approximation), we obtain the analytic bound
\begin{eqnarray}
m_h^2 & \leq & M_Z^2 + (\lambda^2 {\nu}^2 - M_Z^2) \sin^2 2\beta
\nonumber \\ & + & \frac{{\nu}^2}{32\pi ^2} \ln \left(
\frac{M_{SUSY}}{\mu} \right) [(12 h_t^4 - 12{\lambda}^2 h_t^2 -
8{\lambda}^2 k^2 - 24{\lambda}^4) -24 h_t^4 \cos 2\beta \nonumber
\\ & + & (12 h_t^4 + 12{\lambda}^2 h_t^2 + 8{\lambda}^2 k^2 +
8{\lambda}^4) \cos^2 2\beta ],
\end{eqnarray}
where all parameters are evaluated at $M_{SUSY}$. It is easy to
show that this is maximised for $\lambda > {\lambda}_{max}$ and
$k=0$; thus in the full numerical solution of the RG equations
yielding ${\lambda}_i(\mu)$, we set $\lambda = {\lambda}_{max}$
and $k=0$. We take $\mu = 150$ GeV, which turns out to be close
to the bound on $m_h$.

Our procedure, then, is as follows. For a fixed top mass we
determine which values of $h_t(M_{SUSY})$ and
${\lambda}_{max}(M_{SUSY})$ maximise the right--hand--side of
Eq.~(\ref{analyticbound}) by solving the RG equations in order to
calculate the $\delta {\lambda}_i$'s. This process is then
repeated for all values of the top mass required. In table 1 we
present the bound on the lightest CP-even Higgs mass. In row 1 is
the top mass; in row 2 the bound on the lightest CP-even Higgs
mass; in rows 3 and 4, those values of $h_t(M_{SUSY})$ and
${\lambda}_{max}(M_{SUSY})$ which give rise to the bound in row
2. Thus, we find that $m_h \leq 146$ GeV, in reasonable agreement
will similar analyses \cite{rgbound}.

\begin{table}
\caption{\tenrm \baselineskip=12pt Lightest CP-even Higgs mass
bound in the
NMSSM. In row 1 is the top mass, $m_t$, in GeV; in row 2 the
upper bound on the lightest CP-even Higgs mass, in GeV; in rows 3
and 4 those values of $h_t(M_{SUSY})$ and
${\lambda}_{max}(M_{SUSY})$ which produce the bound,
respectively.}
\begin{center}
\begin{tabular}{|c|r|r|r|r|r|r|r|r|r|r|r|}
\hline
\vbox to 12pt {} $m_t$ & 90 & 100 & 110 & 120 & 130 & 140 &
150 & 160 & 170 & 180 & 190 \\
\hline
\vbox to 12pt {} Bound & 146 & 143 & 140 & 137 & 134 & 131 &
128 & 126 & 124 & 123 & 126 \\
\hline
\vbox to 12pt {} $h_t$ & 0.61 & 0.67 & 0.73 & 0.78 & 0.83 &
0.88 & 0.92 & 0.95 & 0.98 & 1.00 & 1.02 \\
\hline
\vbox to 12pt {} ${\lambda}_{max}$ & 0.87 & 0.85 & 0.83 &
0.81 & 0.79 & 0.76 & 0.73 & 0.70 & 0.67 & 0.63 & 0.50 \\
\hline
\end{tabular}
\end{center}
\end{table}

Of course, such a bound is of little experimental interest if it
can never be realised physically. To this end, we have studied
the spectrum of Higgs particles in various regions of parameter
space and determined that, indeed, the bound may become nearly
saturated. We refer the reader elsewhere for further discussion
of these results \cite{elliott}.

To conclude, then, the lightest CP-even Higgs scalar in the NMSSM
must respect the bound $m_h \leq 146$ GeV. This bound is
calculated within the framework of a low energy RG approach
assuming hard decoupling of superpartners and the existence of a
SUSY desert between $M_{SUSY}$ and $M_{GUT}$. The effects of
squarks have been neglected. At present, we are in the process of
calculating the effects of their contributions to the bound, with
preliminary results indicating that they may shift the bound
upwards by only a few GeV for small $m_t$, but by perhaps 20 GeV
for large $m_t$ \cite{squarks}.

\vglue 0.6cm
{\elevenbf \noindent Acknowledgements \hfil}
\vglue 0.4cm
This work was performed in collaboration with S. King and P.
White, University of Southampton, England. T.E. would like to
thank the Science and Engineering Research Council for the
support of a studentship.
\vglue 0.6cm
{\elevenbf \noindent References \hfil}
\vglue 0.4cm

\vglue 0.2cm

\end{document}